# Comparative study of divacancies in 3C-, 4H- and 6H-SiC

Danial Shafizade,[1] Joel Davidsson,[1] Takeshi Ohshima,[2,3] Nguyen Tien Son,[1] and Ivan G. Ivanov[1,*]
*[1]Department of Physics, Chemistry and Biology, Linköping University, SE-58183 Linköping, Sweden*
*[2]National Institutes for Quantum and Radiological Science and Technology, 1233 Watanuki, Takasaki, Gunma 370-1292, Japan*
*[3]Department of Materials Science, Tohoku University, 6-6-02 Aramaki-Aza, Aoba-ku, Sendai 980-8579, Japan*
*[*]Author to whom correspondence should be addressed: ivaiv28@liu.se*

The divacancy comprising two neighboring vacant sites in the SiC lattice is a promising defect for applications in quantum technology. So far, most work is concerned with the divacancy in 4H-SiC, whereas the divacancies in 6H- and 3C-SiC have received much less attention. Here, we outline arguments showing that the neutral charge state of the divacancies in the latter two polytypes is intrinsically stable, in contrast to that in 4H-SiC where the photoluminescence quenches in most materials for certain excitation energies (below approximately 1.3 eV). Divacancies in 6H- and 3C-SiC are anticipated to remain stable for all excitation energies above resonant excitation. We provide new *ab initio* calculation results for the charge transfer levels of divacancies in 6H- and 3C-SiC. We show also that the luminescence from the divacancy in 3C-SiC vanishes with increasing temperature towards room temperature because of the proximity of the excited state to the conduction band.

Silicon carbide hosts several optically active defects possessing spin which contest the properties of the well-known nitrogen-vacancy (NV) center in diamond [1,2]. Apart from the NV center, which also exists in SiC and is isoelectronic with the NV center in diamond, the most studied intrinsic defects in SiC are the silicon vacancy ($V_{Si}$) and the divacancy ($V_CV_{Si}$) [3]. The photoluminescence (PL) signature of the neutral divacancy has been identified first in 4H-SiC by observing the optically detected magnetic resonance (ODMR) of all four divacancy configurations and comparing the observed zero-field splitting of the ground states with theoretical calculations [4]. Similar to the NV center in diamond, the divacancy possesses spin $S = 1$ that can be initialized and read out optically [5], and which shows spin coherence times [6] comparable to those of the NV center in diamond [7]. A significant advantage compared to the NV center in diamond is the infrared emission of the divacancy in SiC, ~ 1.1 eV where the attenuation in optical fibers is an order of magnitude lower than that for the diamond centers, making the divacancies in SiC attractive for longer range quantum communications [2,3].

Most of the studies on divacancy so far have been dealing with 4H-SiC, likely due to the technological maturity of this polytype compared to the other two common polytypes, 6H- and 3C-SiC. Nevertheless, the PL of the neutral divacancy is identified in both 6H-SiC [8,9] and 3C-SiC [10] and it has been demonstrated that the spin coherence properties are similar across the three polytypes [8]. The cubic 3C-SiC displays a single zero-phonon line (ZPL), while the number of ZPLs in the 4H- and 6H-polytypes corresponds to the number of inequivalent configurations, four in 4H- and six in 6H-SiC.

It has been demonstrated that under excitation with photon energies below a certain threshold the divacancy PL in most 4H-SiC specimens quenches [11-13]. This quenching is a challenge in the realization of stable divacancy-based qubits in 4H-SiC [14]. The quenching phenomenon has been interpreted in terms of charge-state switching, albeit different mechanisms for the photoionization process have been considered. Refs. [11-12,15] consider two-photon ionization of the defect into the positive [11,15] or negative [12] charge state. In Ref. [13], we show that the two-photon processes can be neglected at the power densities considered and under non-resonant excitation. The proposed quenching mechanism is accumulation of divacancies in the negative (dark) charge state owing to the capture of free electrons photoionized from other impurities (e.g., shallow N donors) or defects (e.g., carbon vacancies, $V_C$) by the exciting laser. Indeed, recent first-principles calculations show that the (0|–) charge transfer level lies within the bandgap of 4H-SiC, approximately 1.17 – 1.31 eV below the conduction band edge (CBE) for the different divacancy configurations [13]. These values are in good agreement with the experimental threshold values of 1.28 – 1.32 eV quoted in the same work. If the laser excitation energy is below the threshold value for a divacancy configuration, the corresponding PL spectrum quenches because the laser energy is not sufficient for re-ionizing the negatively charged state (denoted $VV^-$) to the neutral one $VV^0$, and the corresponding divacancy configuration accumulates in the $VV^-$ state. A second laser at higher energy, called a repumping laser, is usually used to ionize $VV^-$ and recover the "bright" $VV^0$ state.

We notice that the probability of capturing a free electron by a neutral divacancy is much lower than that by an ionized impurity because the latter have giant capture cross sections associated with the Coulomb center [16,17]. Thus, the electron capture by neutral divacancies is a slow process. In contrast, the re-ionization process is instant when the negatively charged divacancy absorbs a photon of sufficient energy. Consequently, the power required from the repump laser to recover and maintain stable PL from the neutral charge state in an ensemble is much lower than that of the excitation laser, often by several orders of magnitude as confirmed experimentally (cf., e.g., [13]).

We notice further that stable emission from divacancies in 4H-SiC with a single-laser excitation (without a repump laser) can be observed in two cases. The first case is trivial: choose the excitation laser energy above the threshold energy of the divacancy configuration of interest. For instance, laser with energy larger than ~1.32 eV (wavelength shorter than ~939 nm) will excite all divacancy configurations without quenching because the photon energy is sufficient to re-ionize divacancies, which occasionally capture electrons. The second case when stable divacancy PL is possible relates to samples with proper Fermi level position. More precisely, if the Fermi level is at the middle or in the lower half of the bandgap, then there exist no electron traps with electrons in the upper half of the bandgap, and no free electrons are generated by laser energies below the thresholds (in the range below 1.28 – 1.32 eV) [14]. Consequently, the free-electron concentration remains negligible, and the divacancies are in the "bright" neutral charge state without the need of a repumping laser. Such a high-purity semi-insulating (HPSI) 4H-SiC sample is demonstrated in [13,14]. Excitation with photon energies below ~1.24 eV does not lead to PL quenching because there are no electron traps in the upper half of the bandgap to emit free electrons, and the Fermi level is shown by electron paramagnetic resonance to be pinned at the positive charge state of $V_C$, ~1.65 eV below the CBE [13]. However, usually the HPSI 4H-SiC exhibits quenching of the divacancy PL, hence, a repumping laser is required to stabilize the charge state. Alternatively, laser excitation with energy above ~1.32 eV may be used but the photoluminescence excitation (PLE) data of Ref. [13] suggests that this excitation is not optimal, since it reduces the excitation efficiency by a factor of ~3 compared to excitation at ~1.2 eV (see also Fig. 3).

Considering that divacancies in 4H-SiC accumulate in a negatively charged state that cannot be re-ionized to a neutral state with excitation energies below approximately 1.24 eV, unless a repumping laser is used, it is reasonable to speculate about similar conditions in the other two common polytypes of silicon carbide: 3C-SiC and 6H-SiC. First, we consider the (+|0) charge transfer level (CTL) representing the position of the ground state of the neutral charge state of the divacancy in the bandgap.

Theoretical data on the (+|0) CTLs in 3C-SiC and 4H-SiC can be found in Ref. [18]. The calculation for 4H-SiC is repeated in [13] with improved accuracy using a 576-atom supercell (Ref. [18] uses 96-atom one) and yields values for the (+|0) levels of the different divacancy configurations in the range 1.01 – 1.08 eV. These values can be compared with the generic value of 0.92 eV of Ref. [18] obtained from their figure representing the formation energies vs. Fermi level position. Thus, the smaller supercell used in [18] seems to underestimate the (+|0) position with respect to the valence band edge (VBE). The theoretical value for the (+|0) level in 3C-SiC using a 216-atom supercell can also be obtained from the same figure in Ref. [18], 1.10 eV.

Since, to our knowledge, data for the CTLs in 6H-SiC is absent in literature and the accuracy of the data for 3C-SiC may be improved by choosing a larger supercell, we carry out *ab initio* calculations for these two polytypes (calculation details are given further below). We shall also consider the rapid quenching of the divacancy PL in 3C-SiC with increasing temperature compared to that in 4H- and 6H-SiC, a consequence of the (+|0) CTL position and the smallest bandgap of 3C-SiC.

Concerning the (0|–) CTL representing the negative charge state of the divacancy in 3C-SiC, Ref. [18] finds it degenerate with the conduction band (CB), while in our calculation it is within a few meV from the calculated CBE. For 4H-SiC, the calculated (0|–) CTLs in [13] for the four inequivalent configurations agree within 0.1 eV with the experimental values determined in the same work and correctly reproduce the ordering of the levels of the four inequivalent configurations. The generic value of [18] for the position of the (0|–) CTL in 4H-SiC (2.07 eV above the VBM, or 1.22 eV below the CBE) agrees well with the values calculated in [13] which are in the range 1.174 – 1.307 eV below the CBE for the four different configurations. The experimental values from the latter work fall in the range of 1.281 – 1.321 eV for the four inequivalent configurations and are in good agreement with the theoretical values from both works [13,18].

To construct the energy diagram presented in Fig. 1, we also need the values of the low-temperature *electronic* bandgap in the three polytypes. Values for the *excitonic* bandgap can be found in the work by Choyke [19]: 2.390, 3.023, and 3.265 eV for 3C-, 6H-, and 4H-SiC, respectively, accurate to within 1 meV. The free-exciton binding energy added to the excitonic bandgap gives the value of the electronic bandgap. Based on the theoretical calculation of the free exciton binding energy of 27.6 meV and the experimental excitonic bandgap, the value of the electronic bandgap of 3C-SiC is given as 2.416 eV [20]. The free exciton binding energies of 4H- and 6H-SiC have been determined experimentally as 20 meV [21,22] and 60 meV [23], yielding values of the bandgaps of 3.285 and 3.083 eV for 4H- and 6H-SiC, respectively.

We use the available data to build the energy diagram presented in Fig. 1. Bearing in mind that for a given polytype the scatter between the values of the CTLs due to inequivalent configurations for the same charge state is less than ~0.1 eV, we can represent in a diagram each CTLs for this polytype with a generic energy level. The shaded area around the level represents the actual scatter of the levels depending on the configuration when such information is available. This is used in Fig. 1 for the (+|0) and (0|–) CTL manifolds corresponding to the four inequivalent divacancy configurations in 4H-SiC. Here, we use the experimental data for the (0|–) levels and the theoretical data for the (+|0) levels [13]. For the 6H polytype, we calculate the (+|0) and the (0|–) CTLs for the *hh* configuration of the divacancy in

this work and use this as a generic value representing all six configurations (see the calculation details below). Since the exact values for all configurations are not essential for our scope, for 6H-SiC we assume the same spread due to the six inequivalent configurations as in 4H-SiC. 3C-SiC has no inequivalent divacancy configurations, so only a single theoretical level from our work is presented in Fig. 1 for the (+|0) CTL. Since the negative charge state of the divacancy in 3C-SiC (i.e., the (0|–) CTL) is likely degenerate with the CB, it is not present in the diagram in Fig. 1

Here we summarize the calculation details. The CTLs are calculated with density functional theory (DFT) via the Vienna Ab initio Simulation Package (VASP) [24,25] with projector augmented-wave method [26,27]. The volume of the supercell is relaxed with the semi-local exchange-correlation functional by Perdew, Burke, and Ernzerhof (PBE) [28]. For 3C-SiC, the supercell contains 1000 atoms (5x5x5) with the relaxed PBE lattice constant of 4.378 Å. For 6H-SiC, the supercell contains 1536 atoms (8x8x2) with the relaxed PBE lattice constants of $a$ =3.094 Å and $c$ = 15.185 Å, same as used in Ref [9]. For the charge correction $E_{corr}$ , we use Lany-Zunger correction [29] with the dielectric constant of 9.6. The corrections are $0.063*q^2$ eV for 3C and $0.055*q^2$ eV for 6H-SiC.

In the 6H-SiC polytype, we calculate the *hh* configuration of the divacancy, and in both polytypes three charge states (0, ±1) are considered. In each charge state, the ion positions are relaxed using the PBE functional with ionic and electronic stopping parameters $5\cdot10^{-5}$ eV and $1\cdot10^{-6}$ eV, respectively. Then, a single-shot electronic convergence is performed with the HSE06 hybrid functional by Heyd, Scuseria, and Ernzerhof (HSE) [30] and electronic stopping parameter of $1\cdot10^{-2}$ eV on top of the PBE geometries. The cutoff energy of the plane-wave basis set is 400 eV.

The charge transition levels are calculated with

$$\epsilon(q_1,q_2)=\frac{E_{D,q_1}+E_{corr}(q_1)-E_{D,q_2}-E_{corr}(q_2)}{q_2-q_1}\,, \qquad (1)$$

using the HSE energies. For 3C-SiC, the (+|0) level is at 1.181 eV and the (0|−) level at 2.243 eV above the VBE. The latter is within the error margin from the calculated band gap of 2.25 eV, hence, in principal agreement with the result of Ref. [18] finding the (0|−) level degenerate with the CB. For 6H-SiC, the (+|0) level is calculated at 1.119 eV and the (0|−) level at 2.127 eV above the VBE, with calculated band gap of 2.94 eV.

Excitation with photons with energy 1.15 eV is illustrated in Fig. 1 and exemplifies the different behavior of the divacancy PL in 4H-SiC, on the one hand, and in 3C- and 6H-SiC, on the other hand. This excitation is close to resonant excitation in all three polytypes owing to the proximity of their ZPL energies. The energy diagram in Fig. 1 shows that the luminescence in 4H-SiC will be unstable with this excitation and will decay if electron traps within less than 1.15 eV from the CB are ionized. Free electrons will be provided from such traps under the action of 1.15 eV excitation, and some of these will be captured by divacancies and change their neutral charge (bright) to negative (dark). However, since the negative charge state is separated from the CB by more than 1.15 eV (between 1.281 and 1.321 eV for the different configurations [13]) the negative charge state cannot be re-ionized back to neutral without the aid of an additional pump laser at energy > 1.3 eV. On the other hand, the negative charge state of the divacancy [the (0|–) CTL] in 6H-SiC is only ~0.93 eV below the CBE and can be promptly re-ionized to the neutral charge state with resonant or near-resonant excitation (~1.15 eV). Bearing in mind that electron capture from a neutral divacancy is a slow process, while re-ionization of the (0|–) CTL is fast, we expect that no significant accumulation of divacancies in the negative charge state occurs. In other words, the excitation laser will also serve as repumping laser for any other excitation energy above the energy of the corresponding ZPL. A similar scenario is valid for 3C-SiC whether the negative charge is within the bandgap very close to the CBE [our result] or degenerate with the CB [18]. Hence, no quenching of the divacancy PL in any of the configurations is expected in 3C- and 6H-SiC regardless of the Fermi level position. This is confirmed experimentally within the whole range of tuning of the Ti-sapphire laser (up to 1020 nm, or down to ~1.21 eV) and for a few wavelengths close to resonance, as shown in Fig. 3. None of the ZPLs in 3C- and 6H-SiC quenches upon the excitations presented in Fig. 3. The figure displays also the decreasing efficiency of the excitation both far away from the ZPL and with getting too close to it, which can be understood as due to decreasing overlap with the phonon sideband (PSB) mirrored towards higher energies with respect to the ZPL. We notice here that we do not consider resonant excitation; in this case a different PL-quenching mechanism can be envisaged leading to to the ionization of the divacancies to the positive charge state. Consideration of the associated physical mechanism is beyond the scope of the present work and will be undertaken elsewhere.

We notice that the energy positions of the (+|0) CTLs of the divacancies with respect to the VBE in 3C-, 4H- and 6H-SiC are approximately aligned (to within ~0.1 eV), and the ZPLs in the three polytypes are also in the same range (1.09 – 1.13 eV). We observe from the diagram in Fig. 1 that the excited state of the divacancy in 3C-SiC is not far (within ~0.1 eV) from the CBE and the divacancy may become ionized at elevated temperatures (to the positive charge state) due to phonon-assisted ionization of the electron from the excited state to the CB. The process is depicted schematically in the inset of Fig. 2(d) and is analogous to the phonon-assisted ionization of the negative charge state of the divacancy in 4H-SiC at higher temperatures reported earlier [31]. Thus, at elevated temperatures one may expect an abrupt decay of the intensity in the ZPL of the divacancy in 3C-SiC, compared to the ZPL intensities in 4H- and 6H-SiC.

We measure the temperature dependence of the divacancy photoluminescence in 3C-SiC and compare it with that in 4H- and 6H-SiC in Fig. 2. It can be noted that not only the ZPL but also the entire PSB in 3C-SiC vanishes at a temperature above ~140 K whereas the PL from the divacancies in 4H- and 6H-SiC persists up to room temperature. This is further illustrated in Fig. 2(d) which compares the integrated ZPL intensities of some of the lines in 6H- and 4H-SiC with that in 3C-SiC. For instance, the fastest quenching with the temperature PL4 ZPL in 4H-SiC vanishes above ~170 K, but the associated PSB remains up to room temperature. To conclude, the PL from the divacancy in 3C-SiC cannot be observed at temperatures higher than ~140 K, in contrast to the divacancy PL in 4H- and 6H-SiC owing to the proximity of the excited state to the CB and phonon assisted ionization. Thus, the data presented in Fig.2 illustrates two different possible mechanisms of the quenching of the ZPLs in each polytype. Firstly, with increasing the temperature the number of transitions associated with phonons increases compared to the number of zero-phonon transitions (giving rise to the ZPL) because higher vibronic states in both the ground state (GS) and the excited state (ES) become populated. Secondly, if the ES is within the range of the phonon energies of the crystal from the CBE (the case of 3C), phonon assisted ionization of the ES is activated and leads to quenching of the total luminescence from the divacancy. From the calculated data for the (+|0) CTL of 3C-SiC and the experimental ZPL energy of 1.121 eV one estimates that the ES of the divacancy is just about 100 meV from the CBE which is in the range of the phonon energies, i.e., phonon ionization becomes possible at elevated temperature. The temperature dependence in Fig. 2 is obtained with laser excitation of 960 nm (1.29 eV). Since this energy and higher energies used in Fig. 3 exceed the gap between the (+|0) CTL and the CBM in 3C-SiC, we need to comment on why such excitations do not lead to the ionization of the divacancy. Our notion is that the electron is excited in the CB but does not remain free. In fact, thermal equilibrium will be established on time scales much faster than the lifetime of the electron in the excited state. Hence, at low temperature the electron will relax to the lowest available state which is the excited state of the divacancy. Subsequent recombination to the ground state leads to emission of PL photon. Conversely, at higher temperature when significant part of the electrons excited from divacancies remain at energies above the CBE owing to thermal excitation, the probability for migrating to other defects and recombining there (radiatively or non-radiatively) increases, which leads to ionization of the divacancies, decreasing and ultimately vanishing of the divacancy PL with increasing temperature. This qualitative picture is in good agreement with the experimental observations.

In conclusion, our study assesses the stability of the divacancy PL in the three polytypes considered. We find that while the divacancy PL in 4H-SiC quenches in most materials for excitation energies below ~1.3 eV, no such quenching is expected or observed in 6H- and 3C-SiC independent of the Fermi level position due to the smaller bandgaps of these polytypes. Combining theoretical calculations and experimental results, we also find that the neutral charge state [the (+|0) CTL] in the three investigated polytypes is approximately at the same energy separation (to within ~0.1 eV) from the VBE for all divacancy configurations. This leads to a small separation between the excited state of the divacancy and the CBE in 3C-SiC owing to the smallest bandgap of this polytype. This separation is of the order of the phonon energies in SiC. Our experimental data shows that phonon-assisted ionization of the divacancy in 3C-SiC leads to complete quenching of its luminescence above approximately 140 K. This circumstance limits the use of the divacancy in 3C-SiC for quantum applications at elevated temperatures.

**Acknowledgments.** Financial support from the Knut and Alice Wallenberg Foundation (KAW 2018.0071) and the European projects under Horizon Europe (QRC-4-ESP, no. 101129663, QUEST, no. 101156088, and QuSPARC, no. 101186889) is acknowledged. N.T.S. and I.G.I. acknowledge support from Vinnova (grant no. 2024-00461). J.D. acknowledges support from the Swedish Research Council grant no. 2022-00276. The computations were enabled by resources provided by the National Academic Infrastructure for Supercomputing in Sweden (NAISS), partially funded by the Swedish Research Council through grant no. 2022-06725.

The data that support the findings of this study are available from the corresponding author upon request.

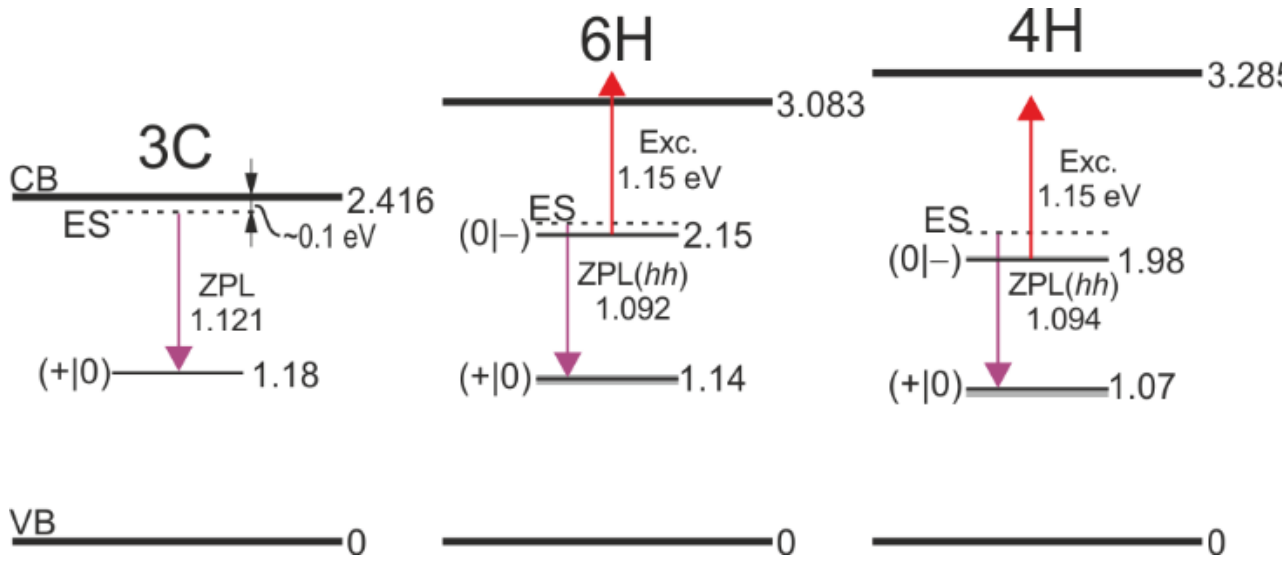

Fig. 1. Energy level diagram for the different charge states of divacancy for the three polytypes of SiC. The (+|0) levels represent the positions of the GS of the neutral charge state and the dashed lines represent the positions of the ES. For 4H- and 6H-SiC, only the positions of the *hh* configurations are shown as an example, but the shaded area around the (+|0) and (0|–) levels indicates the spread of the level due to the other inequivalent configurations. Experimental data for the ZPLs and the CTLs are used whenever available. Thus, the (+|0) levels in 4H-SiC represent the theoretical calculation data but for the (0|–) levels the experimental data of [13] is used. In 6H-SiC only theoretical data for the (+|0) and (0|–) CTLs for the *hh* configuration is available, hence the scatter of the levels is simply assumed the same as in 4H-SiC. For 3C-SiC we use the theoretical data from this work for the (0|+) CTL and the experimental for the ZPL. The upwards vertical arrow (red) illustrates the action of a near-resonant excitation at 1.15 eV, capable of ionizing the negative charge state (0|–) in 6H-SiC, but not in 4H-SiC. All levels and transition energies are labelled with their energy in eV.

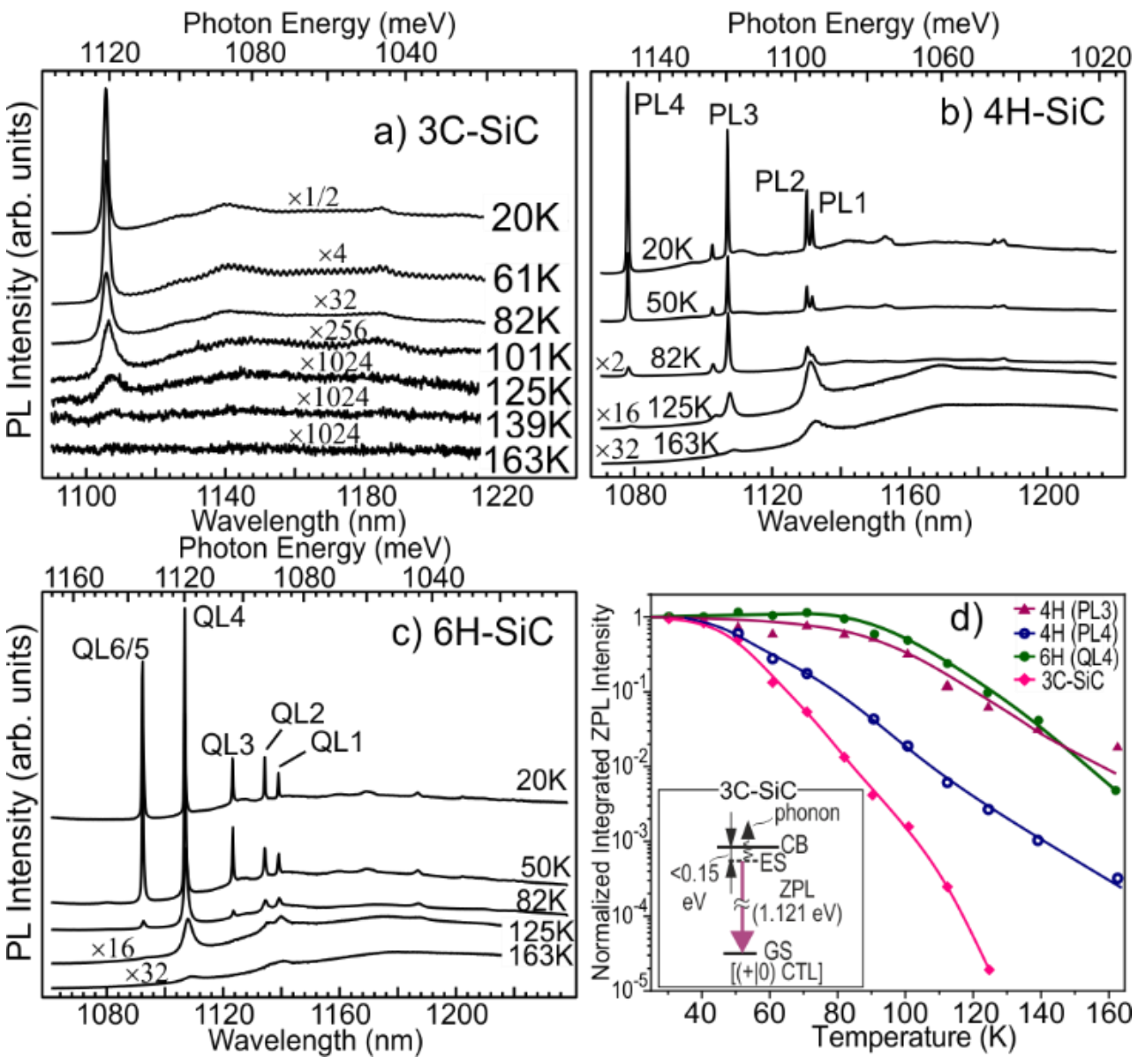

Fig. 2. Temperature dependence of the PL spectra of the divacancy in (a) 3C-SiC, (b) 4H-SiC, and (c) 6H-SiC obtained with excitation at 960 nm. Notice the different scaling factors of the spectra and the complete absence of 3C-SiC divacancy emission at 163 K. (d) Temperature dependence of the integrated intensities of several ZPLs as denoted in the figure. Notice the logarithmic intensity scale. The energy diagram in the inset depicts the phonon assisted ionization from the excited state of the divacancy in 3C-SiC leading to fast PL quenching with temperature.

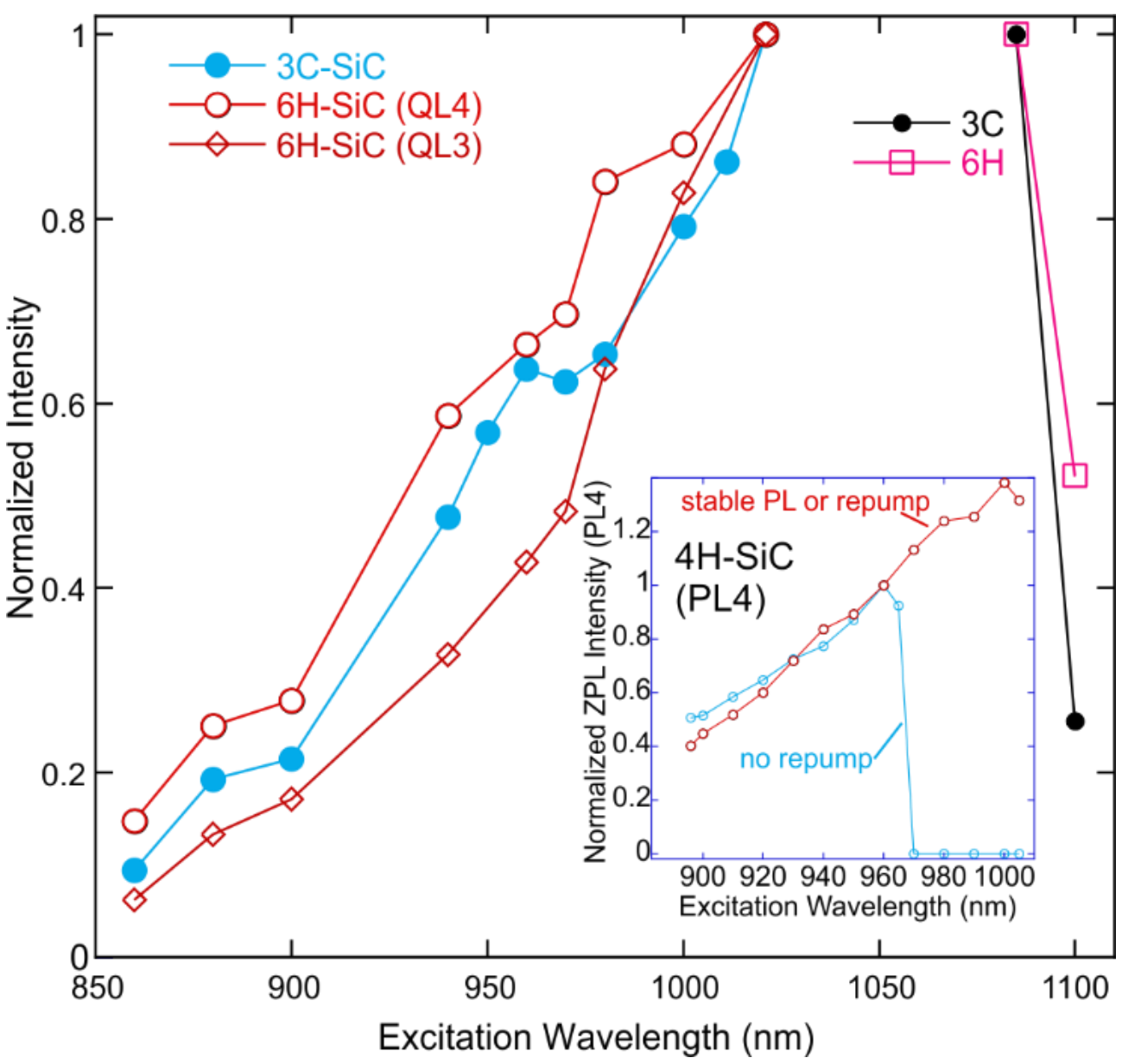

Fig 3. Excitation efficiency dependence on wavelength for the divacancy PL for 3C- and 6H-SiC. The data below 1020 nm is obtained with Ti-sapphire laser keeping the power constant and is normalized at 1020 nm. The scatter of the data points

is largely due to inhomogeneous PL intensity on different points of the samples, thus, the trend for the three ZPLs shown is similar. The data at 1085 and 1100 nm is also obtained by keeping the same power on a Toptica laser and is normalized at 1085 nm. The data from the two lasers cannot be compared directly due to different beam shapes leading to different focusing conditions through the objective, hence the separate normalization. No quenching is observed for the displayed excitation wavelengths. For comparison, the inset shows the data for PL4 in 4H-SiC adapted from [13]. The PL quenches above ~970 nm (full symbols), unless repump laser or sample with suitable Fermi level is used (empty symbols). The trend is similar to that in 3C- and 6H-SiC.